\documentclass[11pt]{article}
\usepackage{graphicx}

\newcommand{\BABARPubYear}    {01}

\newcommand{\BABARProcNumber} {11}
\newcommand{\SLACPubNumber} {8830}

\usepackage{relsize}
\def\babar{\mbox{\slshape B\kern-0.1em{\smaller A}\kern-0.1em
    B\kern-0.1em{\smaller A\kern-0.2em R}}}
\mathchardef\Upsilon="7107

\long\def\inst#1{\par\nobreak\kern 4pt\nobreak
    {\it #1}\par\vskip 10pt plus 3pt minus 3pt}

\begin{document}
{\pagestyle{empty}

\begin{flushright}
SLAC-PUB-\SLACPubNumber \\
\babar-PROC-\BABARPubYear/\BABARProcNumber \\
10 May, 2001 \\
\end{flushright}

\par\vskip 2cm

\begin{center}
\Large \bf Charmless Hadronic $B$ Decays at \babar
\end{center}
\bigskip

\begin{center}
\large 
Thomas Schietinger\\
\vspace{1cm}
Stanford Linear Accelerator Center, Stanford, CA 94309 \\
E-mail: Thomas.Schietinger@SLAC.Stanford.edu\\
\vspace{1cm}
(for the \babar\ Collaboration)
\end{center}
\bigskip \bigskip

\begin{center}
\large \bf Abstract
\end{center}
We present several searches for charmless hadronic two-body
and three-body decays of $B$ mesons from electron-positron
annihilation data collected by the \babar\ detector near
the $\Upsilon(4S)$ resonance.
We report the preliminary branching fractions
${\cal B}(B^0 \rightarrow \pi^+ \pi^-) = (4.1\pm 1.0\pm 0.7) \times 10^{-6}$, 
${\cal B}(B^0 \rightarrow  K^+  \pi^-) = (16.7\pm 1.6\pm 1.3) \times 10^{-6}$, 
${\cal B}(B^0 \rightarrow \rho^\mp \pi^\pm) = (49\pm 13^{+6}_{-5}) \times 10^{-6}$, 
${\cal B}(B^+ \rightarrow \eta' K^+) = (62\pm 18\pm 8) \times 10^{-6}$, 
and present upper limits for nine other decays.

\vfill
\begin{center}
Contributed to the Proceedings of the\\ 
Lake Louise Winter Institute on Fundamental Interactions,\\ 
18--24 February 2001, Lake Louise, Alberta, Canada
\end{center}

\vspace{1.0cm}
\begin{center}
{\em Stanford Linear Accelerator Center, Stanford University, 
Stanford, CA 94309} \\ \vspace{0.1cm}\hrule\vspace{0.1cm}
Work supported in part by Department of Energy contract DE-AC03-76SF00515.
\end{center}

\newpage

\section{Introduction}

The study of $B$ meson decays into charmless hadronic final states plays
an important role in the understanding of CP violation.
In the Standard Model, all CP-violating phenomena
are a consequence of a single complex phase in the Cabibbo-Kobayashi-Maskawa
(CKM) quark-mixing matrix \cite{ckm}.
The Belle and \babar\ collaborations have presented
results \cite{Hara,Lange} on measurements of CP-violating asymmetries 
in $B$ decays into final states containing charmonium, leading to constraints 
on the angle $\beta$ of the CKM unitarity triangle.  Measurements of the 
rates and CP asymmetries for $B$ decays into the charmless final states 
$\pi\pi$ and $K\pi$ can be used to constrain the angles $\alpha$ and 
$\gamma$ of the unitarity triangle \cite{gamalpha}.

\section{Detector and data}

Here we present new measurements of the branching fractions for
charmless hadronic decays of $B$ mesons in the final 
states\footnote{Charge conjugate states are assumed throughout, except 
where explicitly noted.} $\pi^+\pi^-$ and
$K^+\pi^-$, and an upper limit for $B\to K^+K^-$, which are based on a data 
sample consisting of an integrated luminosity of 20.6 fb$^{-1}$ taken near the
$\Upsilon(4S)$ resonance (``on-resonance''), corresponding to 
$(22.57\pm 0.36)\times 10^6$ $B\overline{B}$ pairs.  
A data sample of 2.61 fb$^{-1}$ taken at a
center-of-mass (CM) energy 40 MeV below the $\Upsilon(4S)$ resonance
(``off-resonance'') is used for continuum background studies.

We also report measurements of branching fractions and upper limits for $B$ 
decays into non-resonant three-body modes and modes containing $K^*$, $\rho$, $\omega$,
and $\eta'$ resonances, which are based on a smaller data sample (see Sec.~\ref{sec:osaka}). 
 
The data were collected with the \babar\ detector at the PEP-II 
$e^+e^-$ collider at the 
Stanford Linear Accelerator Center.
The collider is operated with asymmetric beam energies, producing a boost
($\beta\gamma = 0.56$) of the $\Upsilon(4S)$ along the collision axis ($z$).  For
the analyses described in this Letter, the most significant effect of the
boost relative to symmetric collider experiments is to increase the momentum
range of the two-body $B$ decay products from a narrow distribution centered at
approximately 2.6~GeV/$c$ to a broad, approximately flat distribution 
extending from 1.7 to 4.2~GeV/$c$.

At this conference, the \babar\ detector has been described in detail 
by Jim Panetta \cite{Jim}.
Here we only reemphasize that the identification of tracks
as pions or kaons is based on the Cherenkov angle $\theta_c$
measured by a unique, internally reflecting Cherenkov ring imaging
detector (DIRC).  The $K$--$\pi$ separation varies as a function of 
momentum and is better than 8 standard deviations ($\sigma$) at 
1.7~GeV/$c$ and decreases to 2.5$\sigma$ at 4~GeV/$c$.

\section{Analysis of \boldmath{$B^0\to \pi^+\pi^-, K^+\pi^-, K^+K^-$}}

The selection of hadronic events for this analysis is based 
on track multiplicity and event topology.
To reduce background from non-hadronic events, the ratio of Fox-Wolfram
moments \cite{fox} $H_2/H_0$ is required to be less than 0.95 and
the sphericity \cite{spheric} of the event is required to be greater than
0.01.

All tracks are required to have a polar angle
within the tracking 
fiducial region 0.41 $<$ $\theta$ $<$ 2.54 rad and a $\theta_c$
measurement from the DIRC.  The latter requirement is satisfied by
91\% of the tracks in the described fiducial region.  We require a
minimum number of Cherenkov photons associated with each $\theta_c$
measurement in order to improve the resolution.  The efficiency of
this requirement is 97\% per track.  Tracks with a $\theta_c$ within
3$\sigma$ of the expected value for a proton are rejected.
Electrons are rejected based on specific ionization
($dE/dx$) in the DCH system, shower shape in the EMC, and the ratio of
shower energy to track momentum.

The kinematic constraints provided by
the $\Upsilon(4S)$ initial state and relatively precise knowledge of the
beam energies are exploited to efficiently identify $B$ candidates.  
We define a beam-energy 
substituted mass $m_{\rm ES} = \sqrt{E^2_{\mathrm b}-{\mathbf p}_B^2}$, 
where $E_{\mathrm b} =(s/2 + {\mathbf p}_i
\cdot{\mathbf p}_B)/E_i$, and $\sqrt{s}$ and
$E_i$ are the total energies of the $e^+e^-$ system in the CM
and lab frames,
respectively, and ${\mathbf p}_i$ and ${\mathbf p}_B$ are 
the momentum vectors in
the lab frame of the $e^+e^-$ system and the $B$ candidate, respectively.  
The $m_{\rm ES}$ resolution is dominated by the beam energy spread 
and is approximately 2.5~MeV/$c^2$.  
Candidates are selected in the range $5.2<m_{\rm ES}<5.3$~GeV/$c^2$.

We define an additional kinematic parameter $\Delta E$ as the
difference between the energy of the $B$ candidate and half the energy
of the $e^+e^-$ system, computed in the CM system, where the pion mass
is assumed for all charged $B$ decay products.  The $\Delta E$ distribution
is peaked near zero for modes with no charged kaons and shifted
on average $-$45~MeV ($-$91~MeV) for modes with one (two) kaons, where
the exact separation depends on the laboratory kaon momentum.
The resolution on $\Delta E$ is about 26~MeV.  
Candidates with $|\Delta E| < 0.15$ GeV are accepted.

Detailed Monte Carlo simulation, off-resonance data, and events in
on-resonance $m_{\rm ES}$ and $\Delta E$ sideband regions are used to study
backgrounds.  The contribution due to other $B$-meson decays, both
from $b\to c$ and charmless decays, is found to be negligible.  The
largest source of background is from random combinations of tracks and
neutrals produced in the $e^+e^- \to q\bar{q}$ continuum (where $q=u$,
$d$, $s$ or $c$).  In the CM frame this background typically exhibits
a two-jet structure that can produce two high momentum, nearly
back-to-back particles.  In contrast, the low momentum and
pseudoscalar nature of $B$ mesons in the decay $\Upsilon(4S)\to B\overline{B}$
leads to a more spherically symmetric event.  We exploit this topology difference by
making use of two event-shape quantities.  The variable we considered
that has the
greatest discriminating power is the angle $\theta_S$ 
between the sphericity axes evaluated in the CM frame, of the $B$
candidate
and the remaining tracks and photons in the event.  The distribution
of the absolute value of $\cos\theta_S$ is strongly peaked near $1$ for
continuum events and is approximately uniform for $B\overline{B}$ events.  We
require $|\cos\theta_S| < 0.9$, which rejects 66\% of the background
that remains at this stage of the analysis.

The second quantity used in the analysis is a linear combination of
the nine scalar sums of the momenta of 
all tracks and photons (excluding the $B$ candidate decay products) flowing
into 10$^\circ$ polar angle intervals coaxial around the thrust axis of the $B$
candidate, in the CM frame (Fisher discriminant \cite{fisher} ${\cal F}$).
Monte Carlo samples are used to obtain the values of the coefficients, which
are chosen to maximize the statistical separation between signal and
background events.  No restrictions are placed on ${\cal F}$.  
Instead, it is used as an input variable in a maximum likelihood fit, 
described below.

Signal yields are determined from an unbinned maximum likelihood fit
that uses the following quantities: $m_{\rm ES}$, $\Delta E$, $\cal{F}$, and
$\theta_c$. 
The likelihood for a given candidate $j$
is obtained by summing the product of event yield $n_k$ and
probability ${\cal P}_k$ over all possible signal and background
hypotheses $k$.
The $n_k$ are determined by maximizing the extended likelihood
function $\cal L$:
\begin{equation}
{\cal L}= \exp\left(-\sum_{k=1}^M n_k\right)\,
\prod_{j=1}^N \left[\sum_{k=1}^M n_k {\cal P}_k\left(\vec{x}_j;
\vec{\alpha}_k\right)
\right]\, .
\end{equation}
The probabilities ${\cal P}_k(\vec{x}_j;\vec{\alpha}_k)$ are evaluated
as the product of probability density functions (PDFs) for each of the
independent variables $\vec{x}_j$, given the set of parameters
$\vec{\alpha}_k$.  Monte Carlo simulated data is used to validate the
assumption that the fit variables are uncorrelated.  The exponential
factor in $\cal L$ accounts for Poisson fluctuations in the total
number of observed events $N$.  For the $K^\pm\pi^\mp$ terms
the yields are rewritten in terms of the sum $n_{f}+n_{\bar{f}}$ and the asymmetry
${\cal A} =(n_{\bar{f}}-n_{f})/(n_{\bar{f}}+n_{f})$, where
$n_f\,(n_{\bar{f}})$ is the fitted number of events in the mode $B \to
f\,({\overline{B}\to\bar{f}})$.

The parameters for both signal and background $m_{\rm ES}$, $\Delta E$, and
$\cal F$ PDFs are determined from data, and are cross-checked with the
parameters derived from Monte Carlo simulation.  The $\theta_c$ PDFs
are derived from kaon and pion tracks in the momentum range of
interest from approximately 42\,000 $D^{*+}\to D^0\pi^+$ ($D^0\to
K^-\pi^+$) decays.  This control sample is used to parameterize the
$\theta_c$ resolution as a function of track
polar angle.

The results of the fit are summarized in Table \ref{tab:brresults}.
For the decay $B^0\to K^+K^-$ we measure
the branching fraction  ${\cal B} = (0.85^{+0.81}_{-0.66}\pm
0.37)\times 10^{-6}$.
The 90\% confidence level upper limit for this mode is computed 
as the value $n_k^0$ for which
$\int_0^{n_k^0} {\cal L}_{\rm max}\,dn_k/\int_0^\infty 
{\cal L}_{\rm max}\,dn_k = 0.90$, where ${\cal L}_{\rm max}$ is 
the likelihood as a function of $n_k$,
maximized with respect to the remaining fit parameters.  
The result is then increased by the total systematic error.  The reconstruction
efficiency is reduced by its systematic uncertainty in calculating the
branching fraction upper limit.  The statistical significance of a given
channel is determined by fixing the yield to zero, repeating the fit,
and recording the square root of the change in $-2\ln{\cal L}$.

\begin{table*}[!htb]
\begin{center}
\caption{
Summary of results for reconstruction efficiencies
($\varepsilon$), fitted signal yields ($N_S$), statistical significances, and
measured branching fractions (${\cal B}$).
The total number of events entering the ML fit is 16032. 
For the $K^+K^-$ mode the 90\% confidence level (CL) upper limit for the branching
fraction is quoted.  
Equal branching fractions for $\Upsilon(4S)\to B^0 \overline{B}\mbox{}^0$ and $B^+B^-$ are assumed.}
\label{tab:brresults}
\begin{tabular}{lcccc} 
\hline\hline
\vspace{-3mm}&&&&\\
Decay mode  & $\varepsilon$ (\%) & $N_S$ & ~~~Stat. Sig. ($\sigma$)~~~ & 
$\cal B$(10$^{-6}$) \\ 
\vspace{-3mm}&&&&\\
\hline
\vspace{-3mm}&&&&\\
$B^0 \to \pi^+\pi^-$ &  $45$  & $41\pm 10\pm 7$ & 
$4.7$  & $4.1\pm 1.0\pm 0.7$ \\
$B^0 \to K^+\pi^-$ &  $45$ & $169\pm 17\pm 13$ &
$15.8$ & $16.7\pm 1.6\pm 1.3$ \\
$B^0 \to K^+ K^-$  & $43$ & $8.2^{+7.8}_{-6.4}\pm 3.5$  &
$1.3$  & $<2.5$ \\
\hline\hline
\end{tabular}
\end{center}
\end{table*}

Event-counting analyses, based on the same variable set $x_j$ 
as used in the fits, serve as cross-checks for the ML fit results. 
The variable ranges are generally chosen to be tighter
in order to optimize the signal-to-background ratio, or upper
limit, for the expected branching fractions. 
We count events in a rectangular signal region in the 
$\Delta E$--$m_{\rm ES}$ plane, and estimate the background 
from a sideband area.
The branching fractions measured using this technique are in 
good agreement with those arising from the ML fit analysis.
Figure~\ref{fig:cutncount} shows the distributions in $m_{\rm ES}$ 
for events passing the tighter selection criteria of the event-counting
analyses.

\begin{figure}[!tbh]
\begin{center}
\includegraphics[width=9cm]{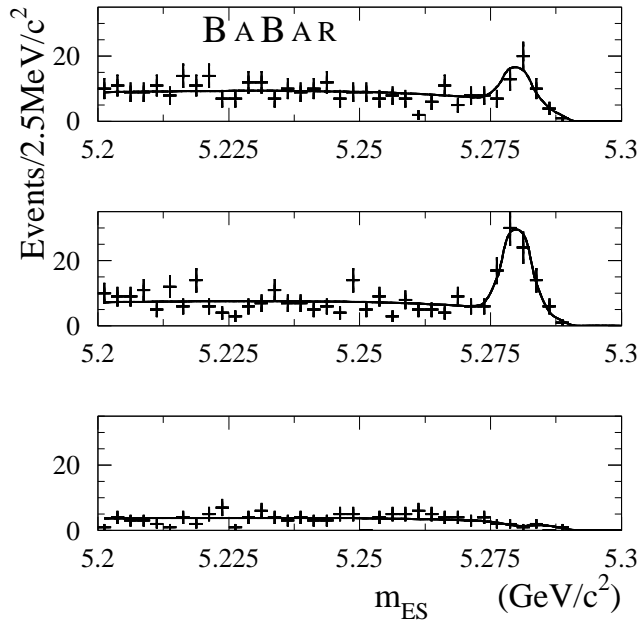}
\caption{The $m_{\rm ES}$ distributions for candidates passing the selection
criteria of the event-counting analyses for the 
$\pi^+\pi^-$ (top), $K^+\pi^-$ (center), and $K^+K^-$ (bottom) mode.}
\label{fig:cutncount}
\end{center}
\end{figure}

The following sources of systematic uncertainty have been considered:
imperfect knowledge of the PDF shapes, which translates into systematic
uncertainties in both the fit yields and asymmetries; systematic uncertainties
in the detection efficiencies, which affect only the branching ratio
measurements; and possible charge biases in either track reconstruction
or particle identification, which affect only the asymmetries. 

The PDF shapes contribute the largest source of systematic uncertainty.  
Systematics due to uncertainties in the PDF parameterizations
are estimated either by varying the PDF
parameters within $1\sigma$ of their measured uncertainties or by
substituting alternative PDFs from independent control samples, and
recording the variations in the fit results.
The largest systematic uncertainties of this type vary across decay
modes and are in the range 1\%--7\% .  The total systematic
uncertainties in the signal yields due to PDF systematics are given
in Table \ref{tab:brresults}.

The overall systematic uncertainties on the branching fractions 
as given in Table \ref{tab:brresults}
are computed by adding in quadrature the PDF systematics
and the systematic uncertainties on the efficiencies.

\section{Non-resonant three-body modes and quasi-two-body modes}
\label{sec:osaka}

In Table~\ref{tab:osaka} we summarize some earlier results that were
obtained with a smaller datasample of 7.7 fb$^{-1}$ on-resonance, corresponding 
to 8.8 million $B\overline{B}$ pairs, and 1.2 fb$^{-1}$ off-resonance.
For the details of these event-counting analyses we refer to Ref.~9.

\begin{table*}[!htb]
\begin{center}
\caption{
Summary of branching fraction measurements.
Inequality denotes 90\% CL upper limit, including
systematic uncertainties.}
\label{tab:osaka}
\begin{tabular}{lc|lc} 
\hline\hline
\vspace{-3mm}&&&\\
Decay mode  & $\cal B$(10$^{-6}$) & Decay mode  & $\cal B$(10$^{-6}$) \\
\vspace{-3mm}&&&\\
\hline
\vspace{-3mm}&&&\\
$B^+ \to \pi^+\pi^-\pi^+$   & $<$ 22 & 
$B^+ \to \eta' K^+$        & $62 \pm 18 \pm 8$ \\

$B^+ \to \rho^{0}\pi^+$    & $<$ 39 & 
$B^0 \to \eta' K^0$        & $<$ 112 \\

$B^0 \to \rho^\mp\pi^\pm$  & $49 \pm 13^{+6}_{-5}$ & 
$B^+ \to \omega K^+/\pi^+$ & $<$ 24 \\

$B^+ \to K^+\pi^-\pi^+$    & $<$ 54 &
$B^+ \to \omega K^0$       & $<$ 14 \\

$B^+ \to \rho^{0}K^+$      & $<$ 29 & 
$B^+ \to K^{*0}\pi^+$      & $<$ 28 \\

\hline\hline
\end{tabular}
\end{center}
\end{table*}

\section{Conclusion}

In summary, we have measured branching fractions for the rare charmless
decays 
$B^0\to\pi^+\pi^-$, 
$B^0\to K^+\pi^-$, 
$B^0 \to \rho^\mp \pi^\pm$, and
$B^+ \to \eta' K^+$, 
and set upper limits on 
$B^+ \to \pi^+\pi^-\pi^+$,
$B^+ \to \rho^{0}\pi^+$,
$B^+ \to K^+\pi^-\pi^+$,
$B^+ \to \rho^{0}K^+$,
$B^0 \to \eta' K^0$,
$B^+ \to \omega K^+/\pi^+$,
$B^+ \to \omega K^0$, and
$B^+ \to K^{*0}\pi^+$.
Our results are in good agreement with earlier measurements \cite{cleobr}.

\section*{Acknowledgments}

We wish to thank our PEP-II colleagues for their
outstanding efforts in providing us 
with excellent luminosity and machine conditions.
This work is supported by the US Department of Energy
and National Science Foundation, the
Natural Sciences and Engineering Research Council (Canada),
Institute of High Energy Physics (China), the
Commissariat \`a l'Energie Atomique and
Institut National de Physique Nucl\'eaire et de Physique des Particules
(France), the
Bundesministerium f\"ur Bildung und Forschung
(Germany), the
Istituto Nazionale di Fisica Nucleare (Italy),
the Research Council of Norway, the
Ministry of Science and Technology of the Russian Federation, and the
Particle Physics and Astronomy Research Council (United Kingdom).


\begin{thebibliography}{99}

\bibitem{ckm} 
N.~Cabbibo, Phys.\ Rev.\ Lett.\ {\bf 10}, 531 (1963);
M.~Kobayashi and T.~Maskawa, Prog. Theor. Phys. {\bf 49}, 652 (1973).

\bibitem{Hara}
T.~Hara, these proceedings.

\bibitem{Lange}
D.~Lange, these proceedings.

\bibitem{gamalpha}
M.~Gronau and D.~London, Phys.\ Rev.\ Lett.\ {\bf 65}, 3381 (1990);
M.~Gronau, J.L.~Rosner and D.~London, {\it ibid.} {\bf 73}, 21 (1994);
R.~Fleischer, Phys.\ Lett.\ B {\bf 365}, 399 (1996);
R.~Fleischer and T.~Mannel, Phys.\ Rev.\ Lett.\ {\bf 57}, 2752 (1998);
M.~Neubert and J.~Rosner, Phys.\ Lett.\ B {\bf 441}, 403 (1998);
M.~Neubert, J. High Energy Phys.\ {\bf 02}, 014 (1999);
M.~Neubert, Nucl.\ Phys.\ Proc.\ Suppl.\ {\bf 99}, 113 (2001).

\bibitem{Jim}
J.~Panetta, these proceedings.

\bibitem{fox}
G.~C.~Fox and S.~Wolfram, Phys.\ Rev.\ Lett.\ {\bf 41}, 1581 (1978).

\bibitem{spheric}
S.L.~Wu, Phys.\ Rep.\ {\bf 107}, 59 (1984).

\bibitem{fisher} CLEO Collaboration, D.M.~Asner {\it et al.,} Phys.\ Rev.\ D {\bf 53}, 1039 (1996).

\bibitem{Osaka}
\babar\ Collaboration, B.~Aubert {\it et al.,} {\it Measurements of charmless three-body
and quasi-two-body $B$ decays}, \babar-CONF-00/15, submitted to the XXX$^{\rm th}$
International Conference on High Energy Physics, Osaka, Japan.

\bibitem{cleobr} 
CLEO Collaboration, D.~Cronin-Hennessy {\it et al.,} Phys.\ Rev.\ Lett.\ {\bf 85}, 515 (2000);
CLEO Collaboration, S.J.~Richichi {\it et al.,} {\it ibid.} {\bf 85}, 520 (2000);
CLEO Collaboration, C.P.~Jessop {\it et al.,}   {\it ibid.} {\bf 85}, 2881 (2000).

\end{thebibliography}
\end{document}